\documentclass[twocolumn,tighten]{aastex631}

\usepackage{amsmath}

\def\bfnabla{{\mbox{\boldmath $\nabla$}}}

\newcommand\bvp{{\mbox{\boldmath $v$}}}

\newcommand\bP{{\mbox{\boldmath $P$}}}

\def\<{\,\langle\langle}
\def\>{\,\rangle\rangle}

\begin{document}

\title{3D Radiation-Hydrodynamical Simulations of Shadows on Transition Disks}

\correspondingauthor{Shangjia Zhang}
\email{sz3342@columbia.edu}

\author[0000-0002-8537-9114]{Shangjia Zhang}
\altaffiliation{NASA Hubble Fellowship Program (NHFP) Sagan Fellow}
\affiliation{Department of Physics and Astronomy, University of Nevada, Las Vegas, 4505 S. Maryland Pkwy, Las Vegas, NV, 89154, USA}
\affiliation{Nevada Center for Astrophysics, University of Nevada, Las Vegas, Las Vegas, NV 89154, USA}
\affiliation{Department of Astronomy, Columbia University, 538 W. 120th Street, Pupin Hall, New York, NY, 10027, USA}

\author[0000-0003-3616-6822]{Zhaohuan Zhu}
\affiliation{Department of Physics and Astronomy, University of Nevada, Las Vegas, 4505 S. Maryland Pkwy, Las Vegas, NV, 89154, USA}
\affiliation{Nevada Center for Astrophysics, University of Nevada, Las Vegas, Las Vegas, NV 89154, USA}

\begin{abstract}
Shadows are often observed in transition disks, which can result from obscuring by materials closer to the star, such as a misaligned inner disk. While shadows leave apparent darkened emission as observational signatures, they have significant dynamical impact on the disk. We carry out 3D radiation hydrodynamical simulations to study shadows in transition disks and find that the temperature drop due to the shadow acts as an asymmetric driving force, leading to spirals in the cavity. These spirals have zero pattern speed following the fixed shadow. The pitch angle is given by tan$^{-1}$($c_s$/$v_\phi$) (6$^{\circ}$ if $h/r$=0.1). These spirals transport mass through the cavity efficiently, with $\alpha \sim 10^{-2}$ in our simulation. Besides spirals, the cavity edge can also form vortices and flocculent streamers. When present, these features could disturb the shadow-induced spirals. By carrying out Monte Carlo Radiative Transfer simulations, we show that these features resemble those observed in near-infrared scattered light images. In the vertical direction, the vertical gravity is no longer balanced by the pressure gradient alone. Instead, an azimuthal convective acceleration term balances the gravity-pressure difference, leading to azimuthally periodic upward and downward gas motion reaching 10\% of the sound speed, which can be probed by ALMA line observations.

\end{abstract}


\section{Introduction} \label{sec:intro}
Shadows are a common feature of protoplanetary disks observed in scattered light images \citep{benisty22}. These shadows are evident in various systems, including narrow shadows observed in HD 142527 \citep{avenhaus17, hunziker21}, HD 100453 \citep{benisty17}, RX J1604.3-2130 A \citep{pinilla15}, DoAr 44 \citep{avenhaus18}, SU Aur \citep{ginski21}, GG Tau A \citep{keppler20}, CV Cha \citep{ginski24}, HD 135344B \citep{stolker16}, and CQ Tau \citep{uyama20, safonov22}, as well as wide shadows in ZZ Tau IRS \citep{hashimoto24}, TW Hya \citep{debes23}, HD 139614 \citep{muro-arena20}, HD 169142 \citep{bertrang18}, HD 143006 \citep{benisty18}, PDS 66 \citep{wolff16}, and HD 163296 \citep{rich19}. Many of these disks are classified as transition disks \citep{vandermarel23}, characterized by a large inner cavity, often accompanied by unresolved inner disks.

The presence of both inner and outer disks suggests a geometric explanation for the observed shadows: a misalignment between the inner and outer disks causes the inner disk to cast a shadow on the outer disk \citep{marino15}. The shadow's extent is influenced by the degree of mutual inclination: mild inclination results in wide shadows, while a highly inclined configuration produces narrow shadow lanes \citep{facchini18}. Many mechanisms could cause misaligned inner and outer disks, including an inclined planet \citep{zhu19, nealon19}, a misaligned central binary \citep{rabago24}, a late infall \citep{kuffmeier21}, and a stellar flyby \citep{nealon20,smallwood24}. Shadows can also be cast by other obscurations, such as dust clumps \citep{rich19}, magnetospheric accretion onto the star \citep{bouvier99}, infall \citep{kuffmeier21, krieger24}, or even a planet with a circumplanetary disk \citep{montesinos21, muley24}.

Evidence of shadows has also been accumulated from ALMA observations. The drops of dust continuum emission in HD 142527 \citep{casassus15}, DoAr 44 \citep{arce-tord23}, and CQ Tau  \citep{ubeira19, safonov22} are aligned with their shadows. The azimuthal variation of CO emission is also aligned with shadows in RXJ1604.3–2130 A \citep{stadler23}. The azimuthal C/O variations in HD 100546 has been attributed to temperature variations caused by shadows \citep{keyte23}.

Most previous studies focused on how disk structures affect shadow appearance, such as interpreting inner disk geometry \citep[e.g.,][]{marino15}, precession rates from multi-epoch observations \citep{pinilla18, debes23}, and surface density and cooling rates from azimuthal temperature variations \citep{cassasus18}. More dedicated studies post-processed hydrodynamical simulations with Monte Carlo Radiative Transfer codes to produce dust and gas emissions \citep{facchini18, nealon19, ballabio21}. However, shadows can also affect disk dynamics. Shadows lower temperature, and thus pressure, creating a pressure difference between shadowed and unshadowed regions, and providing a persistent asymmetric driving force. \cite{montesinos16, montesinos18, cuello19} performed 2D hydrodynamical simulations considering the dynamic effects of shadows, though with simplified heating/cooling treatments. While we were at the final stage of our paper preparation, \citet{su24} conducted a 2D parameter space study on substructures due to shadows, and \citet{qian24} performed a 3D simulation to study the role of shadows on disk eccentricity, both with simplified heating/cooling.

In this letter, we present the first 3D radiation hydrodynamical simulation to study shadows cast on outer transition disks. We focus on a simple configuration of a non-precessing inner disk perpendicular to an optically thin outer disk, examining the dynamical impact by shadows and discuss observational implications on scattered light morphology and ALMA kinematics of shadowed protoplanetary disks. In a subsequent article, we will present a more complete parameter study on shadows in disks. Section \ref{sec:method} details our methods, Section \ref{sec:results} presents our main findings, and Section \ref{sec:discussion} discusses the observational implications, with our conclusions summarized in Section \ref{sec:conclusion}.

\section{Method} \label{sec:method}

A typical transition disk is composed of an often unresolved inner disk and a resolved outer disk \citep{vandermarel23}. In our model, the hydrodynamics is evolved solely in the outer disk, while the inner disk provides asymmetric attenuation of the incident stellar irradiation. We adopted spherical polar coordinates $\rm ({r, \theta, \phi})$ in simulations, while we calculated the initial conditions of density and temperature using cylindrical coordinates ($\rm {R, Z, \phi})$.

\subsection{Disk Setup}
The outer disk setup is similar to that of \citet{zhang24}, featuring a power law surface density, an inner truncation, and an exponential cutoff. We created a wide cavity size of 160 au to keep the disk optically thin to stellar irradiation and to provide ample space for studying the dynamics within the cavity. The gas surface density is given by:

\begin{align}
    \Sigma_\mathrm{g} =& \Sigma_\mathrm{g,0} \mathrm{(R/R_0)^{-1}} \nonumber\\ \times & \Big[\frac{1}{2}
    \mathrm{tanh}\Big(\frac{\mathrm{R-160\ au}}{\mathrm{20\ au}}\Big) + \frac{1}{2}\Big] \nonumber\\ \times & \ \mathrm{exp(-R/100\ au)},
	\label{eq:surface density}
\end{align}
where $\Sigma_\mathrm{g,0}$ is the gas surface density at a reference radius of $\mathrm{R_0}$ = 40 au. Following \citet{zhu12}, $\Sigma_\mathrm{g,0}$ is set to $3\ \mathrm{g\ cm^{-2}}$. If there were no cavity, the disk mass would be 0.01 M$_\odot$.

The outer disk's rotational axis (vertical direction) aligns with z-axis.  We assumed a vertically isothermal and a power-law radial temperature structure, from which the vertical density and velocity structures can be calculated accordingly. The disk is initially in vertical hydrostatic equilibrium, but these initial conditions will transition to a new equilibrium state according to the stellar irradiation once the simulation starts. More detailed setup of the initial conditions can be found in \citet{zhang24}.

We assumed the inner disk's rotational axis is aligned with the y-axis, perpendicular to the z-axis of the outer disk. The inner disk provides the most attenuation at its midplane, which occurs when $\theta_y = \mathrm{arccos}(y/r) \sim \pi/2$, where $y/r = \mathrm{sin}(\theta)\mathrm{sin}(\phi)$.
Consequently, the shadow lanes are centered at $\phi$ = 0 and $\pi$. The stellar irradiation received by the outer disk is given by: 
\begin{align}
\rm &\mathbf{F}_*(r, \theta, \phi) = \left (  \frac{R_*}{r}\right )^2  \sigma_b T_*^4 e^{-\tau} \nonumber \\
&\times \Big\{1 - \mathrm{A}(t)\mathrm{exp} \Big[-\Big(\theta_y - \frac{\pi}{2}\Big)^4 / \sigma^2\Big]\Big\}\hat{\mathbf{r}},
\label{eq:irradiation}
\end{align}
where the first line represents the ray tracing in the outer disk and the second line represents the attenuation due to the inner disk, following \citet{montesinos16}.
$T_*$ and $R_*$ represent the stellar surface temperature and radius, respectively, for which we adopt solar values. $\rm \sigma_b$ denotes the Stefan-Boltzmann constant. $\tau$ is the optical depth at the optical frequency (peak of the stellar spectrum) in the radial direction. $\sigma$ is the shadow width. $\mathrm{A}(t)$ represents the attenuation amplitude. The amplitude of the shadow is zero between t=0 to $t_\mathrm{relax}$. This relaxation time allows the axisymmetric disk to settle to the equilibrium thermal state from the initial locally isothermal state before introducing shadows. Then from t=$t_\mathrm{relax}$ to $t_\mathrm{relax}$+$t_\mathrm{grow}$, the shadow gradually reaches its full amplitude, where $t_\mathrm{grow}$ is the ramp-up time for this attenuation. In expression, 
\begin{equation}
    \mathrm{A}(t) = \mathrm{A_0}\mathrm{sin}^2\Big(\frac{\pi}{2}\frac{\mathrm{min}\{\mathrm{max}\{0,t-t_\mathrm{relax}\}, t_\mathrm{grow}\}}{t_\mathrm{grow}}\Big). \nonumber
\end{equation}
We adopted $\mathrm{A_0} = 0.9$, which means that the inner disk provides at most $\tau \sim 2$ attenuation.
$t_\mathrm{relax}$ was set to 18  P$_0$ and $t_\mathrm{grow}$ to 10 P$_0$, where P$_0$ ($\approx$ 253 yr) is the orbital period at reference radius R$_0$ (= 40 au). $\sigma = 0.3$, which can be treated as an inner disk with aspect ratio $h/r \sim 0.1$, being optically thick in the radial direction until three gas scale heights above the midplane. In Section \ref{sec:scattered}, we will also present a wider shadow simulation ($\sigma$=0.5) as the dynamical effects are stronger. 

As for opacity, we used the DSHARP composition \citep{birnstiel18} and a power law MRN dust size distribution \citep[$n(a) \propto a^{-3.5}$,][]{mathis77}. The minimum grain size $a_\mathrm{min}$ = 0.1 $\mu$m and maximum grain size $a_\mathrm{max}$ = 1 mm. We assumed that only small grains determine the temperature distribution due to their high opacity at the peak of the stellar spectrum; therefore, we considered grains sized between 0.1 and 1 $\mu$m, which account for f$_\mathrm{s}$=0.02184 of the total dust mass. The mass ratio between all the dust and gas was assumed to be 1/100. The opacity values can be found in \citet{zhang24} Figure 1.

Since the disk is optically thin to stellar irradiation across the whole region, the local cooling time is much shorter than the orbital time, effectively making it locally isothermal. This makes our model convenient to be compared with pure hydrodynamical simulations with prescribed temperature structure. For disks with smaller cavity sizes at our fiducial density profile, we expect the disk to remain optically thin inside the cavity, so the temperature structure would be similar to our case. In the ring, however, the local cooling time would be longer. At such locations, the temperature contrast between shadowed and unshadowed regions would be weaker, and the temperature distribution would become more asymmetric to the shadow center \citep{casassus19, su24}. We will present such models in our follow-up publication.

\subsection{Radiation Hydrodynamics}
We utilized the Athena++ \citep{stone20} implicit radiation module \citep{jiang14, jiang21}, which incorporates angle-dependent radiative transfer equations with implicit solvers to accurately model the disk radiation transport. The module can capture both optically thin and thick regimes, shadowing, and beam crossing accurately. Additionally, we incorporated stellar irradiation using long-characteristic ray tracing as a heating source (Equation \ref{eq:irradiation} and \citealt{zhang24}).

Our 3D simulation has 160 cells logarithmically spaced from 0.54 to 16 times the reference radius (R$_0$ = 40 au, so 21.6 au to 640 au from inner to outer boundaries). The polar direction is divided into 128 cells, covering a range from 0.21 to 2.93 radians ($\sim$80$^\circ$ above and below the midplane). The azimuthal direction has 320 cells spanning from 0 to 2$\pi$. For the hydro boundary conditions, we used modified outflow boundary conditions for the inner, outer, upper, and lower boundaries. This means that if the fluid at the boundary is flowing out of the domain, we copy the quantities to the ghost cells as a typical outflow setup. Otherwise, we assign zero velocity in this direction in the ghost cells to avoid the inflow. As for radiation boundary conditions, light beams can freely transport out of the domain. If the beam points inward the computational domain, the radiation is assumed to have the background temperature (10 K), which is a typical temperature of molecular clouds. We adopted periodic boundary condition in the azimuthal $\phi$-direction. Other setups are the same as \citet{zhang24}.

\subsection{Synthetic Observation Setup}
We used the same DSHARP opacity, dust-to-gas ratio, and small grain fraction for the Monte Carlo Radiative Transfer (MCRT) code RADMC-3D \citep{dullemond12} to produce synthetic observations. For the outer disk, we copied the same grid, density, and temperature values from the Athena++ simulations as RADMC-3D inputs. For the inner disk, we extended the grid with the same logarithmic spacing all the way to 3 $r_\odot$. Then we placed a vertically aligned inner disk with its axis aligned with the y-axis (its midplane aligned with the x-z plane). We assumed the inner disk ranges from 3 $r_\odot$ to 5 au and also follows the surface density profile (Equation \ref{eq:surface density}) but without the inner cavity truncation. The aspect ratio of the inner disk is $h/r = 0.1$ at $R_0 = 40$ au (temperature is 61 K at $R_0$), or $h/r = 0.04$ at 1 au. For simplicity, we also assumed the inner disk is vertically isothermal and the radial temperature follows a power-law with an index of -0.5 \citep[e.g.,][]{dullemond18}. The radial and vertical dust density structures along with the opacity of the inner disk set the width of the shadow. We just focused on one setup as our main goal is to demonstrate the dynamical impacts by the shadow rather than perfectly reproducing observations. For near-IR observations, we generated H-band (1.63 $\mu$m) polarized scattered light images in the face-on configuration. For ALMA kinematic observations, we assumed that the abundance of $^{12}$CO is $10^{-4}$ of the total gas mass everywhere to calculate the emission surface of $^{12}$CO (J=3-2) in the face-on configuration.

\section{Results} \label{sec:results}
\begin{figure*}
\includegraphics[width=\linewidth]{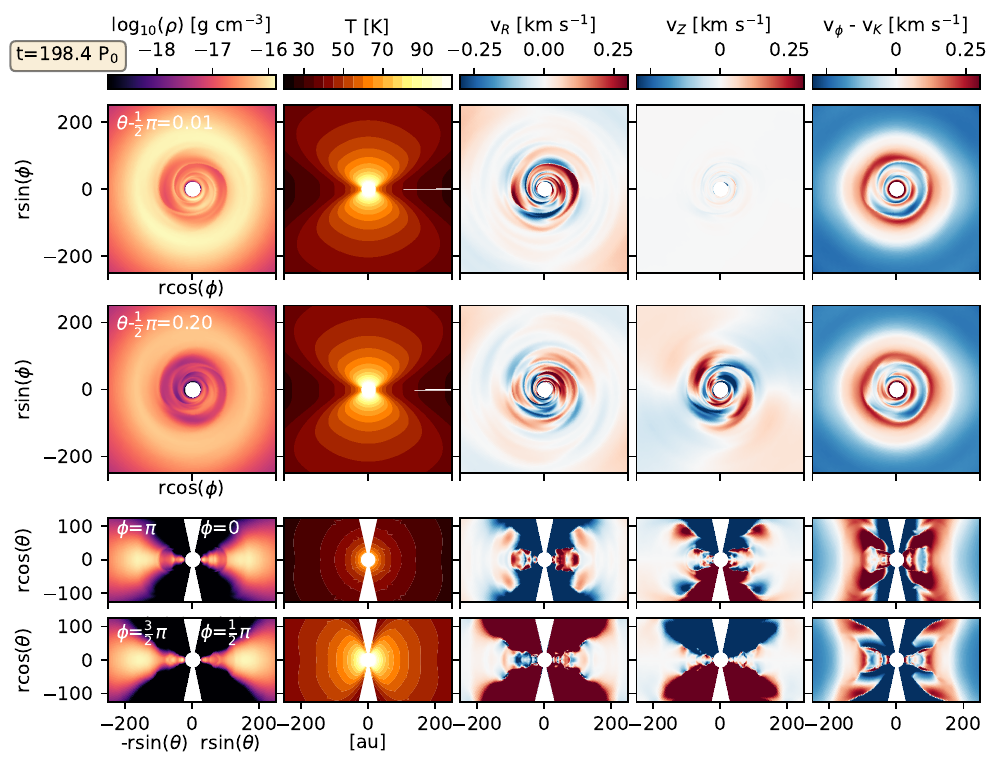}
    \caption{Simulation slices at t = 198.4 P$_0$ (P$_0$: orbital period at 40 au) for various quantities (first row: along the midplane; second row: along 0.2 radians above the midplane; third and fourth rows: vertical slices at $\phi$=0, 0.5$\pi$, $\pi$ and 1.5$\pi$). From left to right, the quantities are density, temperature, radial velocity, vertical velocity, and azimuthal velocity (subtracted by Keplerian velocity). Associated movies in both Cartesian and polar coordinates can be found and downloaded at \url{https://doi.org/10.6084/m9.figshare.26763787.v1}.}
    \label{fig:1}
\end{figure*}

Figure \ref{fig:1} provides an overview of our simulation at a representative time (t = 198.4 P$_0$), showing slices of density, temperature, and velocities. The temperature structure (second column) is established as soon as the attenuation reaches full strength, changing only slightly with time. Since the inner disk is aligned with the x-z plane, the temperatures are lowest at $\phi$=0 and $\pi$ (first and second rows). The temperature is nearly vertically isothermal (third and fourth rows) because the disk is optically thin to stellar irradiation.

The density (first column) in the cavity was low ($\Sigma_g\sim$ $10^{-4}$ g cm$^{-2}$) at the initial condition, but by this time, two trailing spirals connect the cavity edge at 160 au all the way to the inner boundary. The spirals are evident at all layers but exhibit different shapes. At around one to two gas scale heights (second row, $h/r\sim$0.1 inside the cavity and $\sim$0.2 at $\sim$300 au), one spiral could break into two. The ring connected to the inner spirals (between 150-200 au) also show azimuthal density variation. Eventually, two vortices form and merge into one after another 100 P$_0$ (see Figure \ref{fig:1} attached movies and also Figure \ref{fig:2}). The radial velocity $v_R$ (third column) follows the spiral shapes, with some parts of the spirals flowing inward (in blue) and others flowing outward (in red), which is typical for spiral waves. However, at the midplane, the highest density regions of the spirals tend to align with the inflow, consistent with the accumulation of mass in the cavity throughout the evolution.

The vertical velocity ($v_Z$, fourth column) is close to zero at the midplane, indicating that our simulation maintains symmetry across the midplane, and vertical shear instability \citep{nelson13} does not occur due to our low resolution (around five cells per scale height). At around one to two gas scale heights (second row), both the cavity and ring regions show alternating m=2 upward (red) and downward (blue) motions (the direction of $v_Z$ changes sign at each quadrant), which are not related to the spirals but are a unique steady-state solution for a 3D disk with two shadows, as we will detail in Section \ref{sec:vertical}. The azimuthal velocity subtracted by $v_K$ is shown in the fifth column, where the disturbances follow the spirals. The overall azimuthally averaged radial profile of $v_\phi$ follows the pressure structure of a gas ring, with gas orbiting at sub-Keplerian speed (blue) when the pressure gradient is negative (outside the ring and spirals) and super-Keplerian speed (red) when the pressure gradient is positive (inside the ring).

\subsection{Spirals Launched by Shadows}
\begin{figure}
\includegraphics[width=\linewidth]{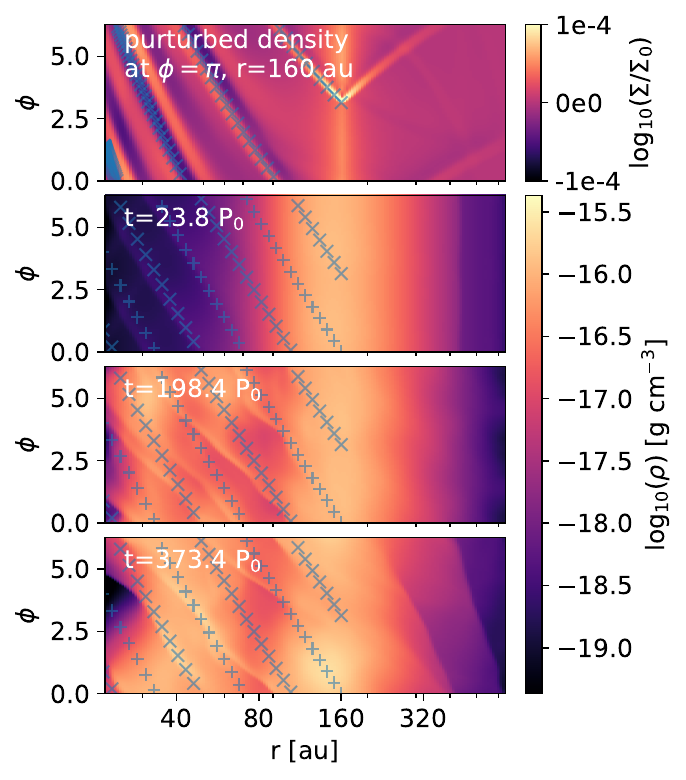}
\caption{$\phi$-ln(r) plots for the surface density of a 2D isothermal simulation (top row) and slices of the midplane density at three different times of the 3D radiation-hydro simulation (bottom three rows). Cross ("$\times$") and plus ("+") signs trace inward-propagating spirals originating from point sources at r=160 au, $\phi$=$\pi$ and 0, respectively, according to the semi-analytical formula from \citet{zhu22}.}
\label{fig:2}
\end{figure}

The launch of the two spirals can be understood as persistent perturbations caused by pressure gradients across the shadows at the cavity edge (one at $\phi$=0 and the other at $\pi$). We can adopt the semi-analytical formula derived by \citet{zhu22} for a point source perturbation, which should apply to a narrow shadow. We take the perturbation's orbital frequency to be zero since the perturbation location is fixed (i.e., the pattern speed is zero in the rest frame), and likewise, the corotation radius to be +$\infty$. Assuming the orbital frequency of the gas disk $\Omega \propto R^{-\alpha_\Omega}$ and sound speed $c_s \propto R^{-\beta}$, the spiral arm should follow:

\begin{align}
\phi = \frac{1}{1-\alpha_\Omega+\beta}\Bigg(\frac{R_{pl}\Omega(R_{pl})}{c_s(R_{pl})} - \frac{R\Omega(R)}{c_s(R)}\Bigg) + \phi_{pl},
\label{eq:spiral}
\end{align}
where $R_{pl}$ and $\phi_{pl}$ are the launching radius and azimuthal angle of the perturbation. The tangent of the pitch angle is simply the ratio between the local sound speed and orbital speed:

\begin{align}
\mathrm{tan}(\psi) = - \frac{c_s(R)}{R\Omega(R)}.
\end{align}

To test this theory, we ran a 2D isothermal simulation ($\alpha_\Omega$=1.5, $\beta$=0), with constant temperature ($h/r$ = 0.1) and surface density ($\Sigma_0$) across all radii, rotating at Keplerian speed. We enforced the density at r=160 au and $\phi$ = $\pi$ to 1.001 $\Sigma_0$ at each time step to produce a persistent perturbation at a point source. Once the simulation started, two spirals were launched and propagated inward and outward through the disk at the local sound speed in the radial direction and the local Keplerian speed in the azimuthal direction. At a steady state (Figure \ref{fig:2}, first row), Equation \ref{eq:spiral} (marked by cross signs) tracks the inward-propagating spiral perfectly. We note that a secondary inner spiral arm occurs around 80 au, which is due to an interference pattern not tracked by this formula \citep{bae17,bae18a,bae18b,miranda19}. We then applied Equation \ref{eq:spiral} to our radiation-hydro simulation and focused on the midplane slice from early to late stages (Figure \ref{fig:2}, second to fourth rows). In the beginning of the linear growth phase (t=23.8 P$_0$) when the shadow is still strengthening, the spiral arms follow Equation \ref{eq:spiral} perfectly. At later stages, vortices form inside the cavity and ring, making the pattern unstable, even though the pitch angle tends to agree with the analytical formula near the launching point. The spirals tend to open up (third row), likely due to the nonlinear shock propagation \citep{goodman01,zhu15}, especially when the density waves travel inwards from the high density to low density region. But from time to time, they realign with the linear phase pitch angle (e.g., t=373.4 P$_0$ inside 60 au).

\begin{figure}
\includegraphics[width=\linewidth]{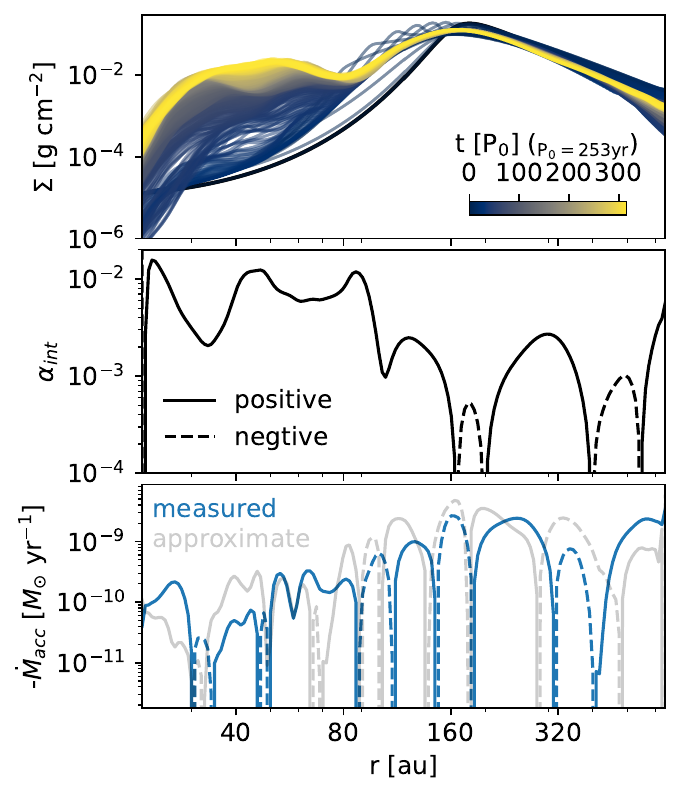}
    \caption{Surface density evolution (top row), vertically integrated $\alpha$ parameter (middle row), and accretion rates (bottom row).}
    \label{fig:2b}
\end{figure}
From Figure \ref{fig:2}, it is evident that the mass in the cavity increases with time, indicating that the spirals drive accretion. We calculated the azimuthally-averaged surface density evolution in Figure \ref{fig:2b} (first row) to confirm this point. As shown in the first row, the surface density at 40 au has increased by two orders of magnitude over 300 P$_0$. To further quantify the accretion by the spirals, we calculated the vertically integrated $\alpha_{R}$ parameter, defined as:
\begin{equation}
\alpha_{int}=\frac{\int T_{R,\phi}dZ}{\int \langle P \rangle_{\phi,t} dZ}\,,\label{eq:alphaint}
\end{equation}
where $T_\mathrm{R, \phi} \equiv \langle \rho v_R v_\phi \rangle_{\phi,t} - \langle v_\phi \rangle_{\phi,t} \langle \rho v_R \rangle_{\phi,t}$. Here, $\langle \rangle_{\phi,t}$ denotes averaging across both $\phi$ and time between 178.56 to 218.24 $P_0$. Note that the underlying assumption for azimuthally averaging is that quantities should be uniform in the background state across $\phi$, which is no longer valid due to the presence of two shadows. Therefore, the values calculated here can at most be considered as an approximation to the transport efficiency. Figure \ref{fig:2b} (second row) shows that $\alpha_{\text{int}}$ is $\sim 10^{-2}$ in the cavity and $\sim 10^{-3}$ at the cavity edge.
We also integrated $\langle \rho v_R \rangle_{\phi,t}$ along the vertical direction to obtain azimuthally-averaged, time-averaged, and vertically integrated radial mass accretion rates ($\dot{M}{_{acc}}= 2\pi R\int \langle \rho v_R \rangle_{\phi,t} dZ$) as functions of $R$. $\dot{M}{_{acc}}$ can also be estimated from the radial gradient of $\alpha{_{int}}$:
\begin{align}
&\dot{M}_{acc}=-\frac{2\pi}{\partial R v_{K}/\partial R}\frac{\partial}{\partial R}\left( R^2 \alpha_{int}\int \langle P \rangle_{\phi,t} dZ\right)\,,\label{eq:mdot}
\end{align}
by integrating the angular momentum equation along $Z$ and assuming that $\langle v_\phi \rangle_{\phi,t}$ equals the midplane Keplerian speed $v_K$.

Figure \ref{fig:2b} (third row) shows the measured and estimated accretion rates. Despite the approximations made in calculating $\alpha_{\text{int}}$ and $\dot{M}_{\text{acc}}$, two curves align quite well. This exercise confirms that the spirals launched by shadows drive accretion at the rate of $10^{-10}$-$10^{-9}$ M$_\odot$ yr$^{-1}$ in our simulation.

\subsection{Vertical Structure \label{sec:vertical}}
\begin{figure}
\includegraphics[width=\linewidth]{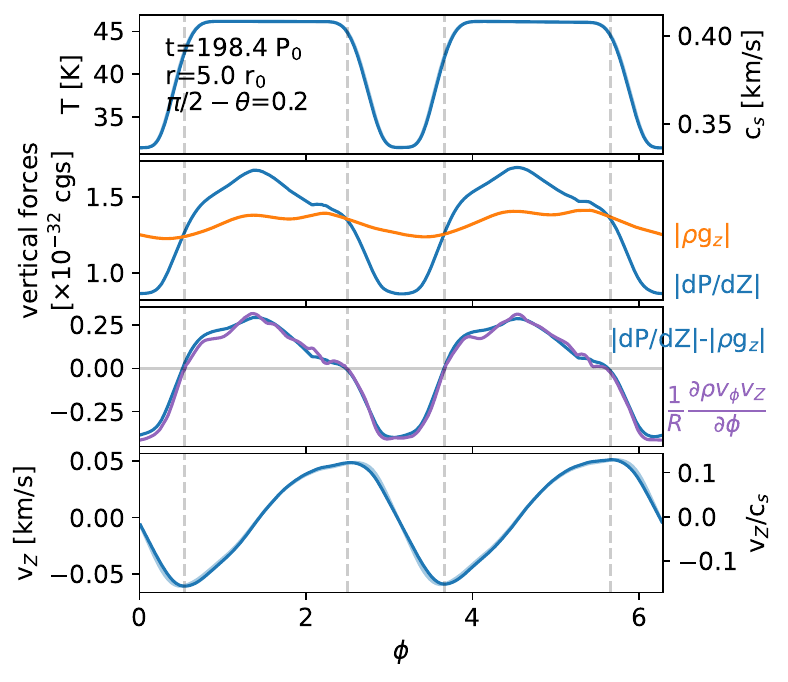}
    \caption{The azimuthal variation of vertical forces (second and third rows), accompanied by temperature/sound speed (first row) and vertical velocity (bottom row) at r=5 $r_0$ (200 au), 0.2 radians above the midplane, t=198.4 P$_0$. The second row shows the vertical pressure gradient in blue and vertical gravity in orange. The third row shows the difference between vertical pressure gradient and gravity in blue, accompanied by the azimuthal derivative of the stress tensor, R$^{-1}$$\partial_\phi(\rho v_\phi v_Z)$, in magenta. The faint blue curves in the first and bottom rows represent quantities associated the sound speed, however they are almost identical to the dark blue curves. Vertical dashed lines are the zeros points in the third panels, or the azimuthal angles where vertical pressure gradient balances the vertical gravity. An associated movie that shows the R-Z slices in the azimuth can be found and downloaded at \url{https://doi.org/10.6084/m9.figshare.26740423.v1}.}
    \label{fig:3}
\end{figure}

With our 3D simulation, we could also study the vertical kinematics due to shadows. In Figure \ref{fig:1}, we have shown that $v_Z$ changes sign in each quadrant. This behavior seems peculiar, as one might expect the gas motion to follow the thermal structure. Specifically, when gas is in shadow, the temperature decreases, leading the gas to collapse toward the midplane, while outside the shadow, the higher temperature would cause the gas to puff up. However, the vertical gas motion does not follow the shadow exactly because the vertical structure does not have enough time to adjust itself. The adjustment takes several sound crossing time or orbital time ($H$/$c_s$ $\sim$ $\Omega^{-1}$), but the shadow only spans a fraction of the orbit.

Instead, the vertical gas motion is still a steady-state feature that can be understood by studying the force balance in the vertical direction using the momentum equation:
\begin{align}
\frac{\partial( \rho\bvp)}{\partial t}+\bfnabla\cdot({\rho \bvp\bvp +{{P\mathbf{I}}}}) &=-\bm{S_r}(\bP) + \rho \mathbf{g},\nonumber\\
\label{eq:hd}
\end{align}
where $\bm{S_r}(\bP)$ is the radiation pressure and $\rho \mathbf{g}$ is the stellar gravity. A classical axisymmetric vertical structure would balance the vertical pressure gradient $\partial P/\partial Z$ and vertical gravity $\rho g_z$. However, by examining all the terms in $Z$-component of the momentum equation, we found that an azimuthal convective accretion term $R^{-1}\partial_\phi (\rho v_\phi v_Z)$ becomes crucial to balance the difference, such that
\begin{align}
   \frac{\partial P}{\partial Z} - \rho g_z  + \frac{1}{R} \frac{\partial (\rho v_\phi v_Z)}{\partial \phi} = 0.
\end{align}

Figure \ref{fig:3} demonstrates the balance of terms along a circle at $r=$200 au and 0.2 radians above the midplane ($\sim$ two gas scale heights in the cavity and one scale height at the outer disk) at t=198.4 P$_0$. The first row shows that the temperature/sound speed drops within the shadows. The second row shows the magnitudes of the vertical gravity (in orange) and pressure gradient (in blue). The vertical gravity remains constant, with variations coming from $\rho$. The pressure gradient varies more and is lower than the gravity term inside the shadow and higher outside the shadow. The vertical lines mark the azimuthal locations where these two terms balance. The blue line in the third row shows the difference between the pressure gradient and gravity, which is almost balanced by the convective acceleration term $R^{-1}\partial_\phi (\rho v_\phi v_Z)$ shown in magenta. Since $v_\phi -v_K$ is much less than $v_K$, the primary contribution to this term is from $v_Z$. That is, when $\partial v_Z$/$\partial \phi$ is positive, $v_Z$ increases and when $\partial v_Z$/$\partial \phi$ is negative, $v_Z$ decreases. The $v_Z$ in the fourth row matches its derivatives perfectly, with the turning points aligning with the zero points in the third row. In fact, our intuition in the beginning of this subsection would be correct if we relate shadow with acceleration, instead of velocity. When a fluid parcel reaches the shadow, it feels a downward acceleration, but the parcel is still moving up so it takes some azimuthal angle for it to move down. Similarly, outside the shadow, the parcel feels an upward acceleration, but since it is moving downward so it takes a certain azimuthal angle for it to move up again. The magnitude of the vertical velocity can reach 10\% of the sound speed at the cavity edge near one gas scale height. This azimuthal variation of $v_Z$ could be a unique feature in ALMA kinematics (Section \ref{sec:kinematics}). We note that a similar discussion on the azimuthal variation on disk scale height due to shadows can be found in Section 5.2 in \citet{benisty17}.

\section{Discussion} \label{sec:discussion}
Given the 3D dynamical consequences the shadows bring to the disk, we expect they will have strong observational implications. We coupled our radiation-hydro simulations with MCRT simulations to produce observational predictions, assuming small grains are well-coupled with the gas. In this section, we provide predictions on near-infrared scattered light images and ALMA line observations.

\subsection{Near-Infrared Scattered Light Images} \label{sec:scattered}
\begin{figure*}
\includegraphics[width=\linewidth]{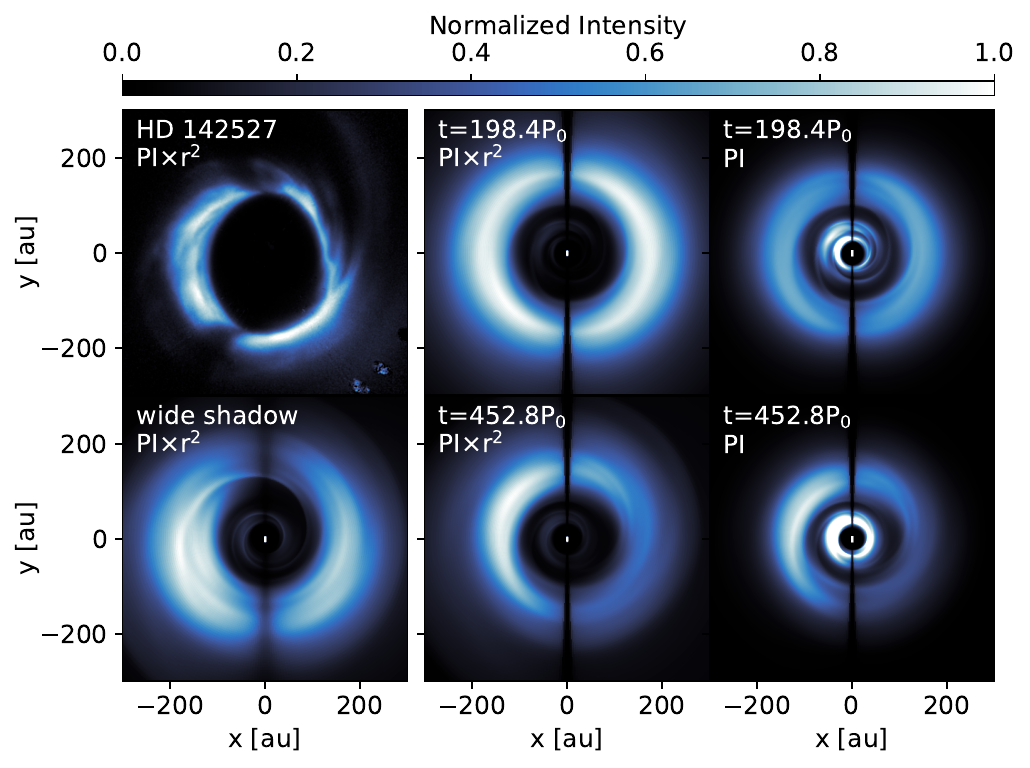}
\caption{Comparison between HD 142527 observation and synthetic observations of radiation-hydro simulations generated by RADMC-3D at H-band (1.63 $\mu$m). Top left: polarization intensity (scaled by square distance from the star) of HD 142527 taken by the IRDIS subinstrument of SPHERE at the VLT \citep{hunziker21}. Bottom left: wide shadow model. Middle panels: fiducial model at t = 198.4 P$_0$ and 452.8 P$_0$. The right two panels show polarization intensity without distance scaling to highlight the inner disk.}
\label{fig:4}
\end{figure*}
The scattered light images of HD 142527 feature two narrow shadow lanes (north and south), spiral arms (east and west), and flocculent streamers on the west side (VLT/SPHERE IRDIS H-band image from \citealt{hunziker21} shown in the top left panel of Figure \ref{fig:4}). From our simulations, we calculated the Polarization Intensity (PI) as $(Q^2 + U^2)^{1/2}$, where $Q$ and $U$ are the Stokes components produced by RADMC-3D, and show them in the rest of Figure \ref{fig:4}. The left four images are scaled by the square distance from the star to highlight the outer disk. The lower left shows a simulation with a wider shadow ($\sigma$ = 0.5). We see spirals on both the east and west sides that are similar to the observation.

The middle panels show our fiducial simulation at 198.4 P$_0$ and 452.8 P$_0$. At the earlier time step, the cavity edge is still close to circular, but at the later time, a vortex forms on the east side, and the west side has at least three flocculent streamers that resemble HD 142527 observation and also GG Tau A \citep{keppler20}. Without scaled by square distance, the right two panels focus on the inner disk where we can clearly see two spiral arms in the cavity.

\subsection{Kinematics in ALMA Line Emissions} \label{sec:kinematics}

\begin{figure*}
\includegraphics[width=\linewidth]{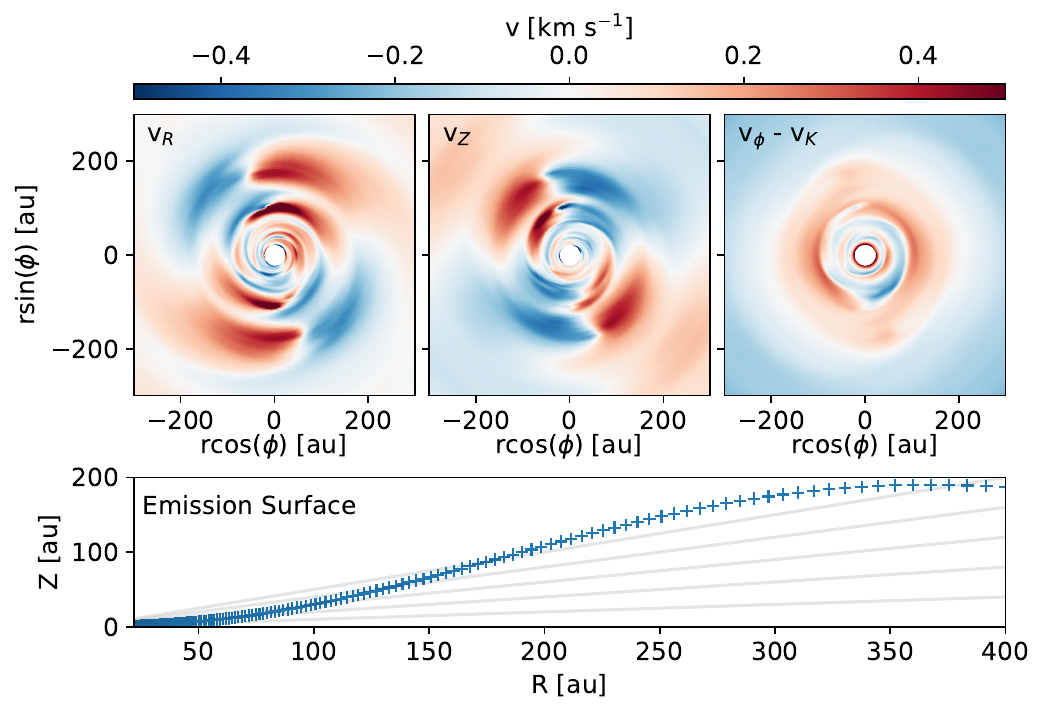}
    \caption{Three velocity components (v$_R$, v$_Z$, and v$_\phi$ - v$_K$) taken at the CO emission surface calculated by RADMC-3D in a face-on view. Slanting lines in the bottom panel represent Z/R=0.1-0.5 in 0.1 spacing.}
    \label{fig:5}
\end{figure*}

We used RADMC-3D to calculate the emission surface of $^{12}$CO(J=3-2) (Figure \ref{fig:5}, bottom row) and measured $v_R$, $v_Z$, and $v_\phi - v_K$ (top row) from the fiducial simulation (t=198.4 P$_0$) at the emission surface. These panels are similar to those in Figure \ref{fig:1}, but they take into account the changing emission surface across the disk. The radial velocity exhibits a strong spiral patterns with significant infall ($\sim$0.5 km s$^{-1}$) inside the cavity. The spiral features are also evident at the cavity edge ($\sim$0.2 km s$^{-1}$). The vertical velocity shows the alternating gas motion, as discussed in Section \ref{sec:vertical}, from the inner cavity to the outer ring. Even at 200 au, the magnitude can reach 0.1 km s$^{-1}$, which can be probed by deep ALMA molecular line observations. These velocity components are all asymmetric, and some have m=2 pattern. Finally, the azimuthal velocity behaves more symmetrically, but the change from sub-Keplerian to super-Keplerian velocities along the azimuthal direction may still be detectable. Additionally, the changing sign of velocity at ($x$,$y$)$\sim$(0,  -100 au) can be mistaken as a Doppler flip for planet-disk interactions \citep{casassus19}. Compared to kinematic features induced by vertical shear instability or planet-disk interactions \citep[e.g.,][]{barraza-alfaro24}, the shadow-induced kinematic features described here operate on a larger scale, making them easier to observe. When the emission surface is close to the midplane, the signature should be sought in $v_R$. As the emission surface moves above the midplane, the signal becomes stronger in $v_Z$. Although this study is limited to one setup, a future parameter space study will systematically quantify these substructures.

Note that the emission surface we calculated here is at the higher end among observations \citep[e.g.,][]{law23}, but it serves as a good example to demonstrate the effect of changing emission surface. The velocity fields at lower emission surfaces can be referenced in Figure \ref{fig:1}, where velocities tend to be lower.

\subsection{Comparison with previous work}
\citet{montesinos16} studied the dynamical consequences of shadows on transition disks using 2D simulations with simplified cooling/heating, finding that spirals can be launched due to the pressure gradients across shadows acting as driving forces. More recently, \citet{su24} extended this work by exploring different shadow widths, strengths, and disk viscosities in full disks without cavities. They found that spirals, rings, and vortices can form depending on the viscosity and shadow strength. Our study differs in two key ways. First, we use 3D full radiation hydrodynamical simulations to accurately model radiation. Second, motivated by observations, we truncate the inner disk in our simulations to better model transition disks. Additionally, we found that the propagation of spirals is highly sensitive to the disk edge, meaning that leaving enough space between the cavity edge and the simulation’s inner boundary is crucial for properly studying disk dynamics. Our study generally confirms previous findings that shadows launch spirals. While \citet{montesinos16} found that spirals form only when stellar irradiation is strong ($L_*$ = 100 $L_\odot$), our fiducial model with just $L_*$ = $L_\odot$ can launch strong spirals inside the cavity. This is consistent with \citet{su24}, who found that spirals can be launched even with small shadow amplitude and weak viscosity. Our 3D radiation hydrodynamical simulations also capture the kinematic structure of the disk, such as the azimuthal modulation of vertical velocity, which could be a unique signature detectable by ALMA line observations. \citet{qian24} used 3D simulations with simplified heating/cooling to study eccentricity excitation in transition disks with one-sided shadows. While a direct comparison is not possible since we focus on two-sided shadows, a giant vortex forms in our simulations at a later stage due to Rossby wave instability \citep{lovelace99}, which may contribute to some eccentricity. In the appendix of \citet{qian24}, two-sided shadows did not develop eccentricity, but the simulation time was an order of magnitude shorter than ours, suggesting that a vortex may eventually develop in an inviscid disk.

\section{Conclusion} \label{sec:conclusion}
We used Athena++ 3D radiation hydrodynamics to study the dynamical effects of shadows cast on transition disks. We focused on the non-precessing inner disk casting shadows on the perpendicular outer transition disk that is optically thin to stellar irradiation. Our findings are as follows:

\begin{itemize}
\item A shadow can act as an asymmetric driving force and launch a spiral. When the perpendicular inner disk casts two shadows in opposite directions, two inward-propagating spirals are launched with zero pattern speed. The pitch angle is given by $\tan^{-1}(c_s/v_\phi)$ and is $\sim 6^{\circ}$ if $h/r = 0.1$.
\item These spirals lead to mass accretion of $\dot{M} \sim 10^{-10}$-$10^{-9}$ M$_\odot$ yr$^{-1}$ within the cavity with $\alpha_{int} \sim 10^{-2}$ in the cavity and $\sim 10^{-3}$ at the cavity edge.
\item Spiral arms, vortices, and flocculent streamers produced by shadows can be seen in the scattered light images which resemble observations. While spirals are the immediate feature caused by shadows, other features can disturb spirals, such as vortices.
\item The shadowed disk has a unique steady-state solution in the vertical direction. In addition to the vertical pressure gradient and gravity, an azimuthal convective acceleration term is needed to balance the vertical momentum equation. This term contributes to the azimuthal variation of vertical velocity, leading to alternating upward and downward gas motions, which can be probed by optically thick ALMA line emissions such as $^{12}$CO. 

\end{itemize}

Future work on exploring different outer and inner disk configurations, such as outer disk cavity size, surface density, shadow width, strength, number of shadows, precession rates, and mutual inclinations, will open up new windows to interpret observations and distinguish them from other substructures formation mechanisms such as instabilities and perturbers.

\section*{Acknowledgement}
We thank the anonymous referee for their constructive review. All simulations are carried out using computers from the NASA High-End Computing (HEC) program through the NASA Advanced Supercomputing (NAS) Division at Ames Research Center. Support for this work was provided by NASA through the NASA Hubble Fellowship grant \#HST-HF2-51568 awarded by the Space Telescope Science Institute, which is operated by the
Association of Universities for Research in Astronomy, Inc., for NASA, under contract NAS5-26555. S.Z. and Z.Z. acknowledge support through the NASA FINESST grant 80NSSC20K1376. Z. Z. acknowledges support from NASA award 80NSSC22K1413.

\vspace{5mm}
\facilities{VLT/SPHERE IRDIS}

\software{Astropy \citep{2013A&A...558A..33A,2018AJ....156..123A}, CMasher \citep{2020JOSS....5.2004V}, Athena++ \citep{stone20}, RADMC-3D \citep{dullemond12}, SciPy \citep{2020SciPy-NMeth}, Matplotlib \citep{hunter07}
          }

\appendix

\end{document}